\documentclass{emulateapj}
\newcommand{\etal}{et\,al.}
\newcommand{\halpha}{H$\alpha$}
\newcommand{\lsim}{\raise0.3ex\hbox{$<$}\kern-0.75em{\lower0.65ex\hbox{$\sim$}}}
\newcommand{\msun}{M$_{\odot}$}

\newcommand{\kms}{km\,s$^{-1}$}
\begin{document}
\slugcomment{Accepted for publication in the Astrophysical Journal Letters}
%-----------------------------------------------------------------------------%
\title{Discovery of a Gas-Rich Companion to the Extremely
  Metal-Poor Galaxy DDO 68}
%-----------------------------------------------------------------------------%

%------------------------------
%Author list - emulateapj style
%------------------------------
\author{
John M. Cannon\altaffilmark{1}, 
Megan Johnson\altaffilmark{2},
Kristen B.W. McQuinn\altaffilmark{3}.
Erik D. Alfvin\altaffilmark{1}, 
Jeremy Bailin\altaffilmark{4},
H. Alyson Ford\altaffilmark{5},
L\'eo Girardi\altaffilmark{6},
Alec S. Hirschauer\altaffilmark{7},
Steven Janowiecki\altaffilmark{7},
John J. Salzer\altaffilmark{7},
Angela Van Sistine\altaffilmark{7},
Andrew Dolphin\altaffilmark{8},
E.C. Elson\altaffilmark{9},
Baerbel Koribalski\altaffilmark{2},
Paola Marigo\altaffilmark{10},
Jessica L. Rosenberg\altaffilmark{11},
Philip Rosenfield\altaffilmark{10},
Evan D. Skillman\altaffilmark{3},
Aparna Venkatesan\altaffilmark{12}
Steven R. Warren\altaffilmark{13}
}

\altaffiltext{1}{Department of Physics \& Astronomy, Macalester College, 
1600 Grand Avenue, Saint Paul, MN 55105, USA; jcannon@macalester.edu}
\altaffiltext{2}{Australia Telescope National Facility, CSIRO Astronomy 
\& Space Science, PO Box 76, NSW, 1710, Epping, Australia}
\altaffiltext{3}{Minnesota Institute for Astrophysics, University of Minnesota,
Minneapolis, MN 55455, USA}
\altaffiltext{4}{Department of Physics and Astronomy, University of Alabama, 
Box 870324, Tuscaloosa, AL 35487-0324, USA}
\altaffiltext{5}{National Radio Astronomy Observatory, P.O. Box 2, Green Bank, WV 24944, USA}
\altaffiltext{6}{Osservatorio Astronomico di Padova-INAF, Vicolo dell'Osservatorio 5, I-35122 Padova, Italy}
\altaffiltext{7}{Department of Astronomy, Indiana University, 727 East
  Third Street, Bloomington, IN 47405, USA}
\altaffiltext{8}{Raytheon Company, 1151 E. Hermans Road, Tucson, AZ 85756, USA}
\altaffiltext{9}{Astrophysics, Cosmology and Gravity Centre (ACGC), Department of Astronomy, University of Cape Town, Private Bag X3, Rondebosch 7701, South Africa}
\altaffiltext{10}{Dipartimento di Fisica e Astronomia Galileo Galilei, Universit{\'a} degli Studi di Padova, Vicolo dell'Osservatorio 3, I-35122 Padova, Italy}
\altaffiltext{11}{School of Physics, Astronomy, and Computational Science, George Mason University, Fairfax, VA 22030, USA}
\altaffiltext{12}{Department of Physics and Astronomy, University of San Francisco, 2130 Fulton Street, San Francisco, CA 94117, USA}
\altaffiltext{13}{Department of Astronomy, University of Maryland, CSS Bldg.,
  Rm. 1024, Stadium Dr., College Park, MD 20742-2421, USA}

%---------------------------
%Author list - journal style 
%---------------------------

%\author{John M. Cannon}
%\affil{Department of Physics \& Astronomy, Macalester College, 1600 Grand 
%Avenue, Saint Paul, MN 55105}
%\email{jcannon@macalester.edu}

%\author{Evan D. Skillman} 
%\affil{Department of Astronomy, University of Minnesota, Minneapolis, MN 55455}
%\email{skillman@astro.umn.edu}

%-----------------------------------------------------------------------------%
\begin{abstract}
%-----------------------------------------------------------------------------%

We present HI spectral-line imaging of the extremely metal-poor galaxy
DDO\,68.  This system has a nebular oxygen abundance of only $\sim$3\%
Z$_{\odot}$, making it one of the most metal-deficient galaxies known
in the local volume.  Surprisingly, DDO\,68 is a relatively massive
and luminous galaxy for its metal content, making it a significant
outlier in the mass-metallicity and luminosity-metallicity
relationships.  The origin of such a low oxygen abundance in DDO\,68
presents a challenge for models of the chemical evolution of
galaxies. One possible solution to this problem is the infall of
pristine neutral gas, potentially initiated during a gravitational
interaction. Using archival HI spectral-line imaging obtained with the
Karl G. Jansky Very Large Array\footnote{The National Radio Astronomy
  Observatory is a facility of the National Science Foundation
  operated under cooperative agreement by Associated Universities,
  Inc.}, we have discovered a previously unknown companion of DDO
68. This low-mass (M$_{\rm HI}$ $=$ 2.8\,$\times$\,10$^{7}$ \msun),
recently star-forming (SFR$_{\rm FUV}$ $=$ 1.4\,$\times$\,10$^{-3}$
M$_{\odot}$\,yr$^{-1}$, SFR$_{\rm H\alpha}$ $<$ 7\,$\times$\,10$^{-5}$
M$_{\odot}$\,yr$^{-1}$) companion has the same systemic velocity as
DDO 68 (V$_{\rm sys}$ $=$ 506 \kms; D $=$ 12.74\,$\pm$\,0.27 Mpc) and
is located at a projected distance of $\sim$42 kpc. New HI maps
obtained with the 100m Robert C. Byrd Green Bank Telescope provide
evidence that DDO 68 and this companion are gravitationally
interacting at the present time. Low surface brightness HI gas forms a
bridge between these objects.

\end{abstract}						

\keywords{galaxies: evolution --- galaxies: dwarf --- galaxies:
irregular --- galaxies: individual (DDO68, DDO 68 C)}

%-----------------------------------------------------------------------------%
\section{Introduction}
\label{S1}
%-----------------------------------------------------------------------------%

The most metal-poor galaxies in the local universe provide important
constraints on models of galaxy evolution.  To date the lowest
measured nebular oxygen abundance in a star-forming galaxy is 3\% of
the Solar value.  Four known galaxies have this abundance value: the
starburst galaxy SBS\,0335$-$052W \citep{izotov05}, the BCD galaxy
I\,Zw\,18 \citep{skillman93}, the dwarf galaxy Leo P
\citep{skillman13}, and the subject of this letter, DDO\,68
\citep{pustilnik05}.

While the extremely low oxygen abundance of DDO\,68 makes it an
interesting system, its other physical characteristics make it a
critical testbed for our understanding of the chemical evolution of
galaxies.  Specifically, DDO\,68 is an outlier on the mass-metallicity
(M-Z) relationship \citep{pustilnik05,berg12}; it is overly massive
compared to the other known systems with comparable metallicity.  The
HI mass of DDO\,68 is four times larger than that of I\,Zw\,18
\citep{vanzee98} and 3 orders of magnitude larger than that of Leo\,P
\citep{ezbc14}.  The deviation of DDO\,68 from the M-Z relation is so
pronounced that it has been excluded from recent works attempting to
calibrate this relationship at the lowest abundances \citep{berg12,skillman13}.

\begin{figure*}
\epsscale{1.0}
\plotone{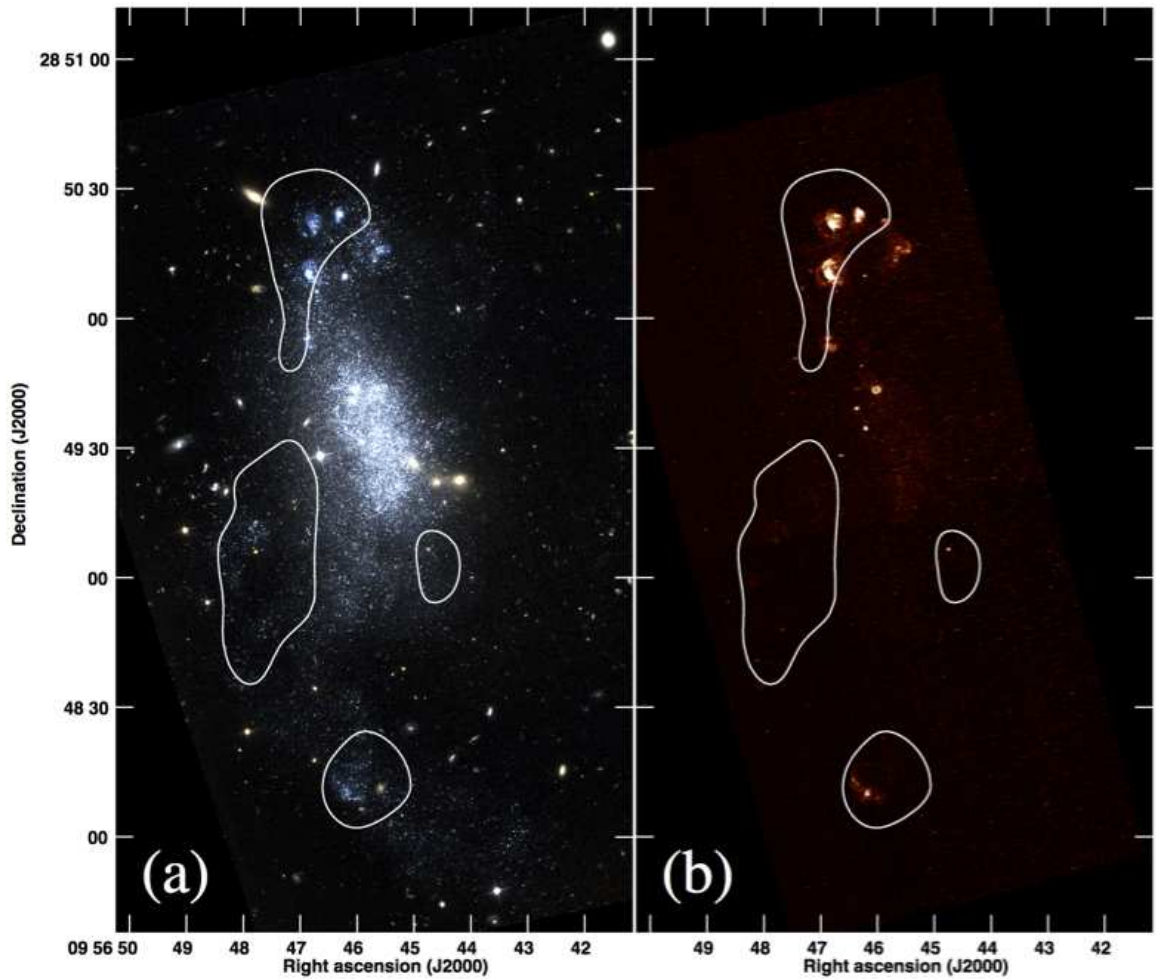}
\epsscale{1.0}
\caption{Color HST/ACS image (a, created using F606W and F814W
  filters) and continuum-subtracted HST \halpha\ image (b) of DDO\,68.
  Overlaid on both panels are the 2\,$\times$\,10$^{21}$ cm$^{-2}$
  column density contours derived from the 15\arcsec\ resolution VLA
  HI images; the same contours are shown in Figure 2.}
\vspace{0.5 cm}
\label{figcap1} 
\end{figure*}

\begin{figure*}
\epsscale{1.0}
\plotone{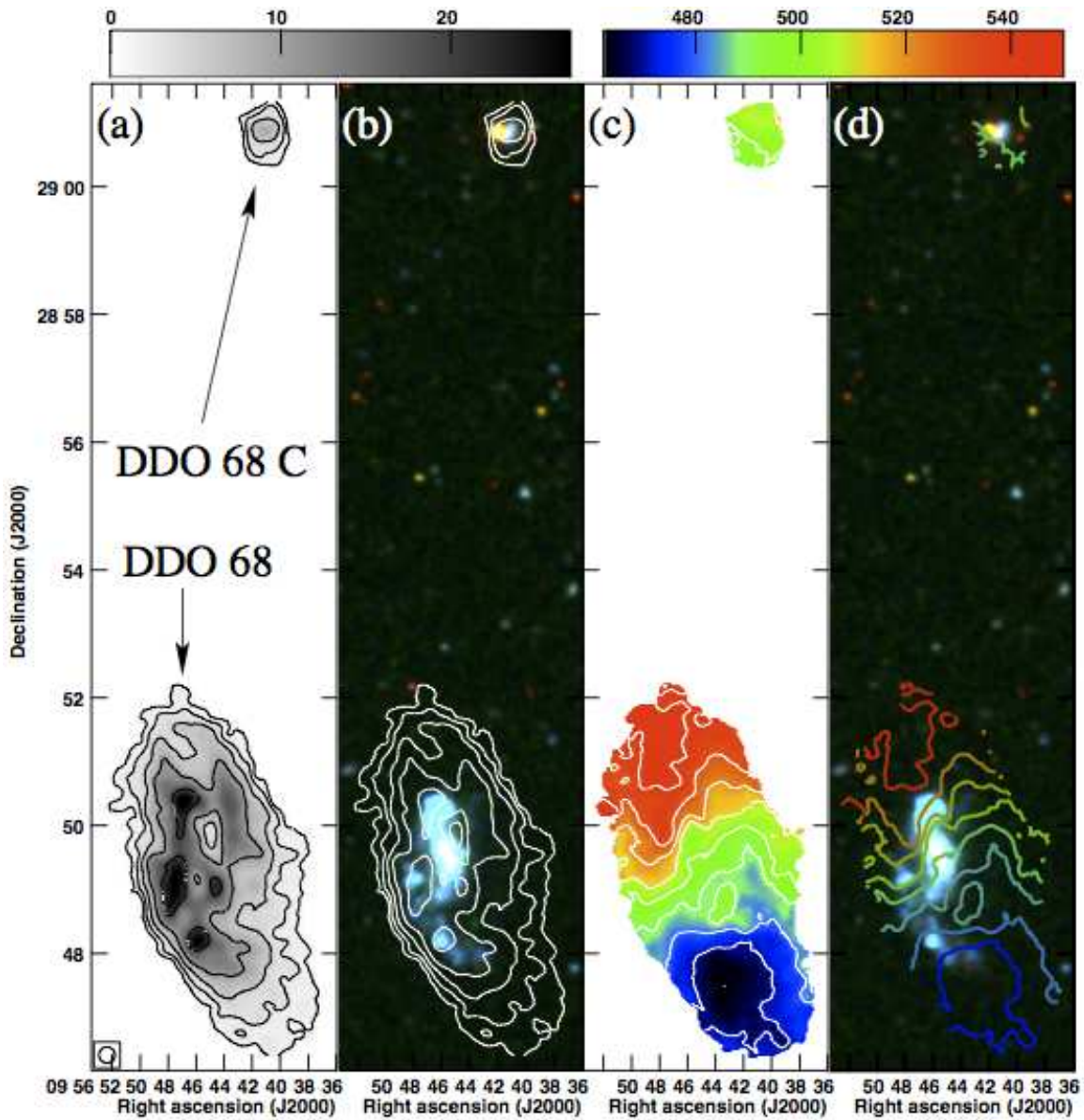}
\epsscale{1.0}
\caption{HI, optical and near-UV images of DDO\,68 and DDO\,68\,C.
  Panel (a) shows the VLA HI column density image (15\arcsec\ beam
  shown at bottom left) overlaid with contours at the
  (1.25,2.5,5,10,20)\,$\times$\,10$^{20}$ cm$^{-2}$ levels; the same
  contours are overlaid on a spatially-smoothed 3-color image in panel
  (b) showing the DSS2-blue, GALEX near-UV, and GALEX far-UV images as
  red, green, and blue, respectively.  Panel (c) shows the intensity
  weighted velocity field of the system, with contours spaced in 10
  \kms\ intervals between 470 and 540 \kms; the same contours are
  overlaid on the 3-color image in panel (d).}
\vspace{0.5 cm}
\label{figcap2}
\end{figure*}

The presence of an extremely metal-poor ISM in a massive dwarf galaxy
poses a serious obstacle for models of the chemical evolution of
galaxies.  One possible origin for these enigmatic properties is the
infall of pristine gas.  As argued in \citet{ekta10}, the infall of
primordial gas may produce lower metallicities and lower effective
chemical yields.  Based on multiple lines of evidence, this scenario
may be at work in DDO\,68.  As the archival Hubble Space Telescope
(HST) image in Figure~\ref{figcap1}(a) shows, the stellar morphology
of DDO\,68 is severely disturbed; an elongated stream of stars extends
southward over an arcminute ($>$3.7 kpc) from the main stellar body.
This stream is especially prominent in the GALEX images shown in
Figure~\ref{figcap2}.  The continuum-subtracted HST H$\alpha$ image
shown in Figure~\ref{figcap1}(b) reveals widespread massive star
formation (SF) in DDO\,68.  Curiously, most of this nebular emission
is concentrated in four SF complexes in the north of the galaxy, and
in one complex in the stellar stream to the south; each of these HII
regions is co-spatial with high column density HI gas (N$_{\rm HI}$
$>$ 2\,$\times$\,10$^{21}$ cm$^{-2}$, or $\sigma$$_{\rm HI}$ $>$ 16
\msun\,pc$^{-2}$) that is significantly displaced from the center of
the main stellar component of DDO\,68.  In agreement with the previous
studies of the HI morphology and dynamics of DDO\,68 by \citet{stil02}
and \citet{ekta08}, we find that the HI morphology of DDO\,68 is
significantly disturbed.  Further, we have discovered a previously
unknown, gas-bearing companion system that is connected to DDO\,68 by
a bridge of low surface brightness HI gas.  Taken together, these
lines of evidence suggest that DDO\,68 is undergoing an interaction or
accretion event; this may support the infall hypothesis.

%-----------------------------------------------------------------------------%
\section{Observations and Data Reduction}
\label{S2}
%-----------------------------------------------------------------------------%

HI spectral-line observations of DDO\,68 were acquired with the Karl
G. Jansky Very Large Array (VLA) in the C configuration in November,
2002 for program AT\,288.  These data divide the 1.5 MHz total
bandwidth into 256 channels, delivering a spectral resolution of 1.19
km\,s$^{-1}$\,ch$^{-1}$.  The primary and phase calibrators were 3C286
and 0958$+$324, respectively.  The total on-source integration time
was approximately 8 hours.  The VLA data were calibrated and imaged
using standard prescriptions in the AIPS
environment\footnote{Developed and maintained by NRAO}.  Residual flux
rescaling was enforced during the image production process
\citep{jorsater95}.  The final beam size is 15\arcsec, and the rms
noise in the final data cube is 0.5 mJy\,Bm$^{-1}$. Moment maps were
derived using the techniques described in \citet{walter08}.

\begin{figure*}
\epsscale{1.2}
\plotone{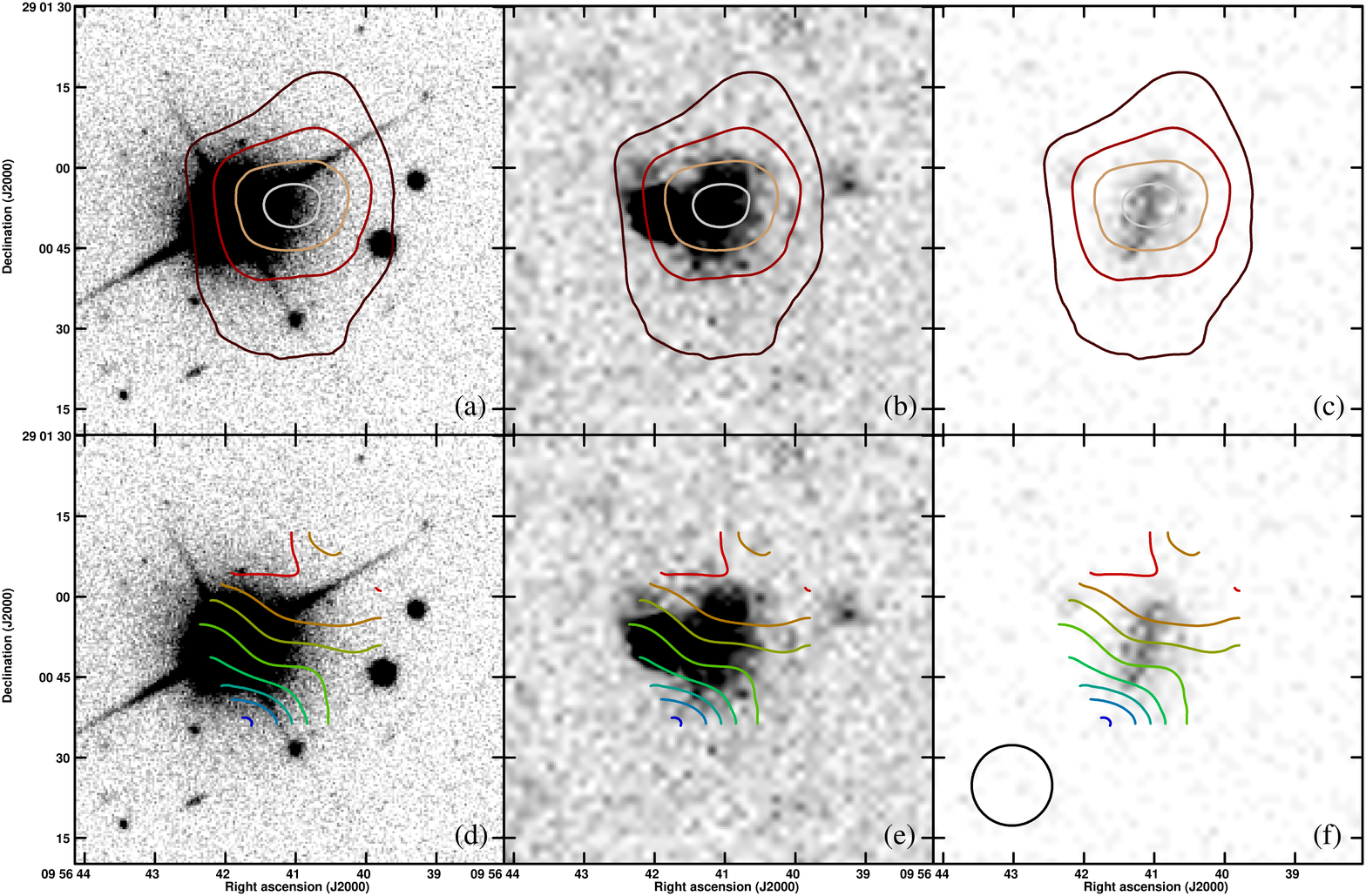}
\epsscale{1.0}
\caption{HI, optical and UV images of the newly-discovered galaxy
  DDO\,68\,C.  Panels (a) and (d) show the SDSS r-band image; panels
  (b) and (e) show the GALEX near-UV image; panels (c) and (f) show
  the GALEX far-UV image.  Overlaid on panels (a), (b), and (c) are
  contours showing HI column densities at the
  (2,4,6,8)\,$\times$\,10$^{20}$ cm$^{-2}$ levels.  Overlaid on panels
  (d), (e), and (f) are isovelocity contours between 494 and 508 \kms,
  in intervals of 2 \kms.  The 15\arcsec\ beam is shown in panel (f).}
\label{figcap3}
\end{figure*}

HI 21-cm spectral-line imaging of the DDO\,68 system was obtained with
the 100m Robert C. Byrd Green Bank Telescope (GBT) in ``on the fly
mapping'' (OTF) mode in January and February, 2014, for programs
AGBT/13B-169 (P.I. McQuinn) and AGBT/13B-459 (P.I. Cannon).  Briefly,
the GBT spectrometer acquired data using in-band frequency switching
with a bandwidth of 12.5 MHz, delivering a spectral resolution of
0.158 km\,s$^{-1}$\,ch$^{-1}$.  The mapping region covered
$\sim$1.1\arcdeg\ and was Nyquist sampled.  Data calibration and
baseline fitting was performed using standard
GBTIDL\footnote{Developed by NRAO; documentation at
  http://gbtidl.sourceforge.net.} routines, with slight modifications
compared to the previous data reduction methods in
\citet{johnson13}. Specifically, after calibrating each row in the
map, all baseline structures were removed by subtracting the average
of the first and last four pixels in each row from the row itself.
This improved fitting method delivers superior baseline fits compared
to a more simplistic low-order polynomial fit to line-free spectral
regions. Calibrated spectra were exported to AIPS for imaging.  All
spectra were combined into a single database and imaged using the
SDGRD task with a spherical Bessel function \citep{mangum07}; the
final, smoothed velocity resolution was 15.46 km\,s$^{-1}$\,ch$^{-1}$.

\begin{figure*}
\epsscale{0.85}
\plotone{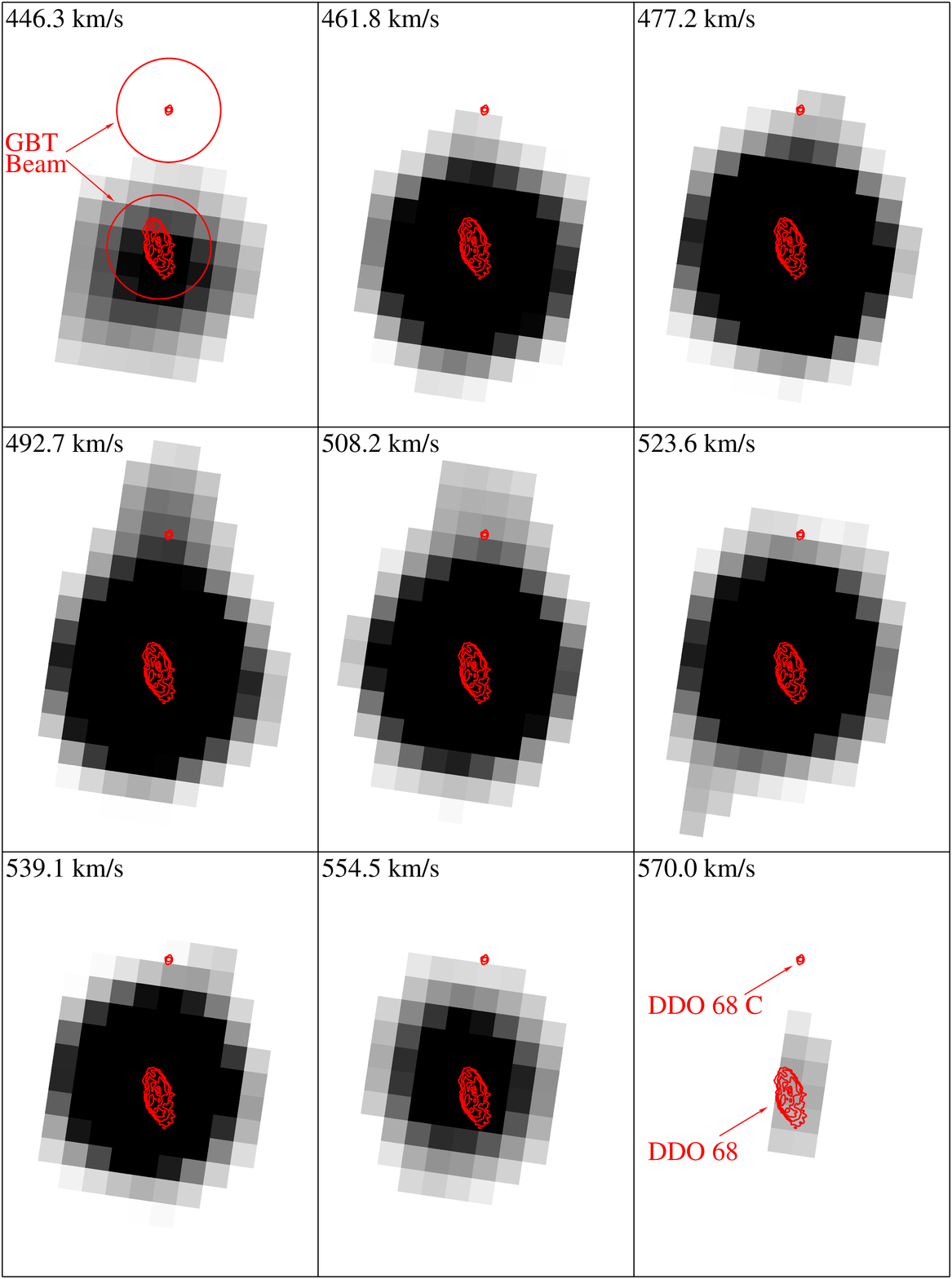}
\epsscale{1.0}
\caption{Channel maps of the spectrally smoothed GBT OTF mapping HI
  cube of the DDO\,68 system; velocities are shown in the upper left
  of each panel.  Pixel intensities range from 0.0 K (white) to 0.075
  K (black); the field of view of each panel is
  $\sim$24.9\arcmin\,$\times$\,33.4\arcmin.  Red contours show the
  same VLA HI column density contours as in Figure 2.  The GBT beam
  size (523.3\arcsec) is shown by the two red circles in the first
  panel; DDO\,68 and DDO\,68\,C are each labeled in the last panel.
  The most extended northern and southern very low column density HI
  gas (V$=$493 and V$=$524 \kms) have velocities in the opposite
  directions as those visible in the adjacent parts of DDO\,68; this
  could provide evidence for dynamically decoupled gas flows along a
  filament or remnant flows from the recent merger discussed in
  \citet{ekta08}.}
\label{figcap4}
\end{figure*}

Optical images of DDO\,68 and the companion system described below
were acquired with the WIYN 0.9m telescope\footnote{The WIYN 0.9m
  telescope is operated by WIYN Inc. on behalf of a Consortium of nine
  partner Universities and Organizations (see
  http://www.noao.edu/0.9m). WIYN is a joint partnership of the
  University of Wisconsin at Madison, Indiana University, Yale
  University, and the National Optical Astronomical Observatory.}  on
26 February, 2014. Two 20-minute exposures in a narrowband
\halpha\ filter, and one 4-minute exposure in a broadband R filter,
were acquired.  Continuum subtraction followed standard prescriptions.

Archival HST, GALEX, and SDSS data were downloaded from the respective
archives.  The HST images of DDO\,68 were acquired in program 11578
(P.I. Aloisi).  Standard photometric analysis of these data were
performed using the DOLPHOT software package \citep{dolphin00}.  The
resulting photometry yields a TRGB distance of 12.74\,$\pm$\,0.27 Mpc,
which is used in this paper.  This value is significantly larger than
previous (indirect method) distance estimates (e.g., {Pustilnik
  \etal\ 2005}\nocite{pustilnik05} estimate D$\simeq$6.5 Mpc).
Accounting for the peculiar motions discussed in \citet{tully08},
\citet{pustilnik11} and \citet{karachentsev13} estimated distances of
$\sim$10 Mpc.  Most recently, \citet{tikhonov14} use the same HST data
as analyzed here to derive a distance of 12.3\,$\pm$\,0.3 Mpc; they
also interpret the color-magnitude diagram and stellar population
distributions as evidence for an ongoing merger, and identify an
interacting companion galaxy (``DDO\,68\,B'') that makes up the bulk
of the stellar stellar stream seen in the HST and GALEX images.

\begin{figure*}
\epsscale{1.2}
\plotone{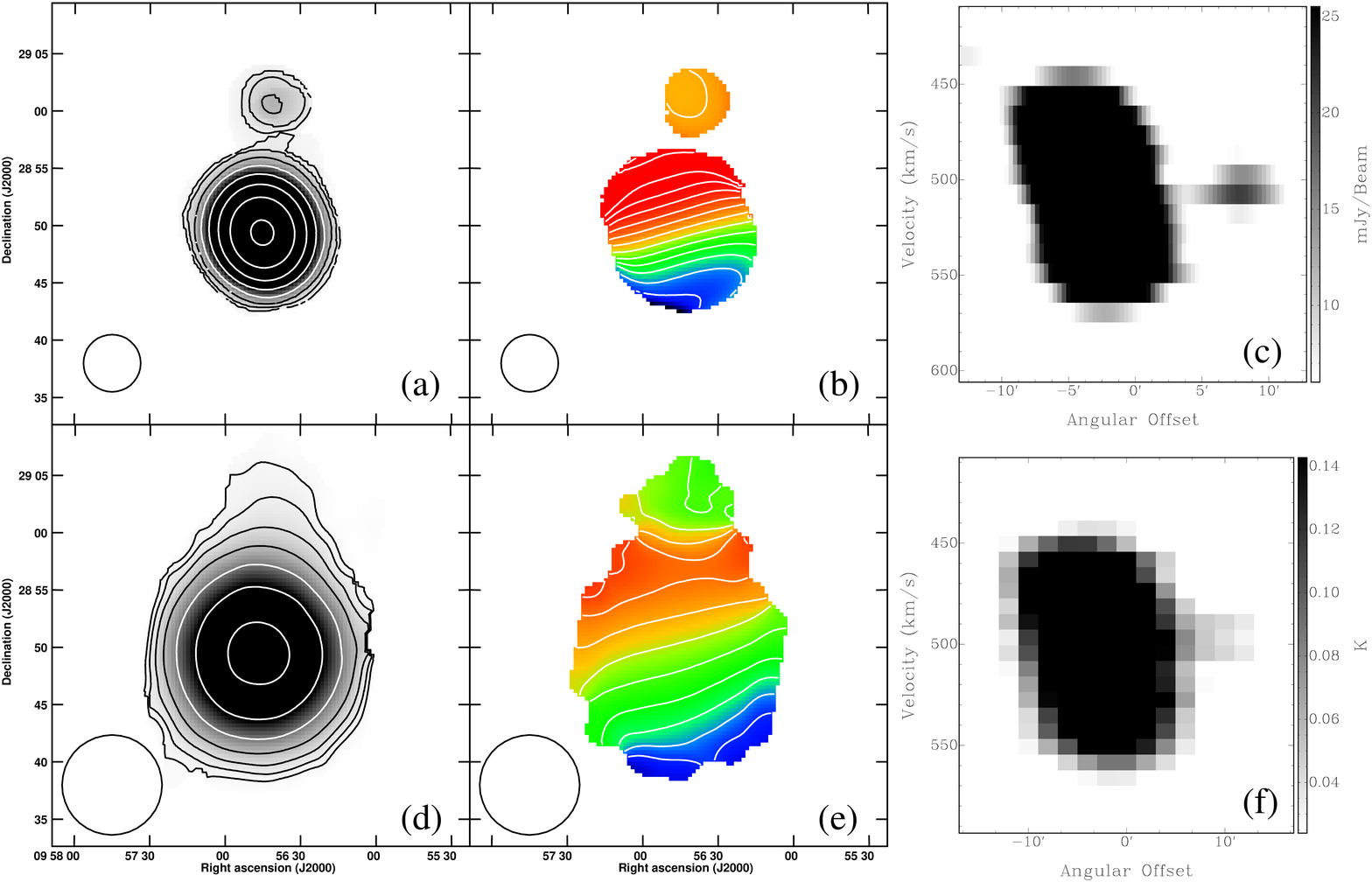}
\epsscale{1.0}
\caption{HI column density images, velocity fields, and PV slices of
  the DDO\,68 -- DDO\,68\,C system.  Panels (a), (b), and (c) show the
  VLA data, tapered to an angular resolution of 300\arcsec, while
  panels (d), (e), and (f) show the GBT data at an angular resolution
  of 523.3\arcsec; beam sizes are shown as circles in the bottom left
  of panels (a), (b), (d) and (e).  The contours in (a) show HI column
  densities of (2,4,8,16,32,128,256)\,$\times$\,10$^{18}$ cm$^{-2}$;
  the contours in (d) show HI column densities of
  (1,2,4,8,16,32,128,256)\,$\times$\,10$^{18}$ cm$^{-2}$.  The
  contours in (b) and (e) show velocities between 475 and 530 \kms, in
  intervals of 5 \kms.  The PV slices in panels (c) and (f) are taken
  at a position angle of 355\arcdeg\ (measured counter-clockwise from
  north) and pass through the HI surface density maxima of both
  DDO\,68 and DDO\,68\,C.}
\label{figcap5}
\end{figure*}

%-----------------------------------------------------------------------------%
\section{The DDO\,68 -- DDO\,68\,C System}
\label{S3}
%-----------------------------------------------------------------------------%

In Figure~\ref{figcap2} we present the HI surface density image and
the intensity weighted HI velocity field of the DDO\,68 system.  The
HI morphology of DDO\,68 is significantly disturbed: the HI surface
density contours are compressed on the eastern side of the galaxy, but
are diffuse and filamentary to the south and west.  The UV emission
shows a similar morphology; the apparent stellar stream that extends
to the south and west is co-spatial with the HI extending in the same
sense.  

The velocity field of DDO\,68 shown in Figure~\ref{figcap2} indicates
coherent rotation. However, significant irregularities in the disk are
apparent (note the kinks in the isovelocity contours in the velocity
range 490-520 \kms).  We thus only estimate the bulk kinematics via
tilted ring analysis, and defer a full mass decomposition until a
later work.  Using the GIPSY task ROTCUR, we derived a representative
rotation curve with the following dynamical parameters: V$_{\rm sys}$
$=$ 505.5 \kms, major axis position angle (measured counterclockwise
from north toward the receding side of the disk) $=$ 19.6\arcdeg,
inclination $i$ $=$ 65.2\arcdeg.  A range of dynamical center
positions yielded similar flat rotational velocities v$_{\rm c}$
$\simeq$45 \kms\ at radii of 180\arcsec\ (11.1 kpc).  The challenges
with fitting the rotation curve, and the estimates of the flat
rotational velocity, are in agreement with the work of \citet{ekta08}.
A simple \begin{math} \frac{V_{c}^2{\cdot}r}{G}\end{math} calculation
gives a total dynamical mass estimate of M$_{\rm dyn}$
$>$5.2\,$\times$\,10$^{9}$ \msun. This value is a lower limit because
it only includes the mass within the HI radius, and because we have
not applied a correction for possible non-circular motions.
The derived HI flux integral (S$_{\rm HI}$ 26.0\,$\pm$2.6 Jy\,\kms)
implies M$_{\rm HI}$ $=$ (1.0\,$\pm$\,0.15)\,$\times$\,10$^{9}$ \msun.  This can be compared
with the stellar mass, derived using the total Spitzer 3.6 $\mu$m and
4.5 $\mu$m fluxes from \citet{dale09} and the formalism presented in
\citet{eskew12}: M$_{\star}$ $\simeq$2\,$\times$\,10$^{8}$ \msun.
Within the HI radius, DDO\,68 is a dark-matter dominated galaxy.

In the VLA data cube we unexpectedly identified a nearby, gas-rich
galaxy that has a systemic velocity identical to that of DDO\,68.  We
hereafter refer to this companion as DDO\,68\,C.  As
Figure~\ref{figcap2} shows, the angular separation of
11.34\arcmin\ (measured from the UV centroid position of DDO\,68\,C at
09$^{\rm h}$56$^{\rm m}$41.07$^{\rm s}$,
$+$29\arcdeg00\arcmin50.74\arcsec\ to the adopted dynamical center
position of DDO\,68 at 09$^{\rm h}$56$^{\rm m}$45.79$^{\rm s}$,
$+$28\arcdeg49\arcmin32.9\arcsec) implies a physical separation of
$\sim$42 kpc.  To our knowledge this source does not appear in the
literature; it is not mentioned in the previous HI studies by
\citet{stil02} and \citet{ekta08}.

DDO\,68\,C is detected at high significance in the HI data. As shown
in the moment zero image presented in Figure~\ref{figcap3}, the HI
column densities peak at 8.8\,$\times$\,10$^{20}$ cm$^{-2}$ at
15\arcsec\ (930 pc) resolution.  Based on the derived flux integral
(S$_{\rm HI} =$ 0.73\,$\pm$0.11 Jy\,\kms), the total neutral hydrogen
mass M$_{\rm HI} =$ (2.8\,$\pm$\,0.5)\,$\times$\,10$^{7}$ \msun.  The
neutral hydrogen reservoir of DDO\,68 is roughly 35 times more massive
than that in DDO\,68\,C.

The lack of previous information about DDO\,68\,C may be due to the
superposition of the Milky Way foreground star TYC\,1967$-$1114$-$1
\cite[m$_{\rm V}$ $=$ 10.8;][]{hog00}. This star has unsaturated
photometry in the Tycho, 2MASS and WISE catalogs; the
solar-metallicity ATLAS9 spectrum \citep{castelli03} that best-fits
this photometry is a T$_{\rm eff}=$5000\,K dwarf with spectral type
between K0V and K2V in the \citet{pickles98} library.  As shown in
Figure~\ref{figcap3}, this star is saturated in the SDSS-r band image.
It is prominent in the GALEX near-UV, and is slightly above the noise
level in the GALEX far-UV.  The latter image clearly delineates the
UV-bright stellar component of DDO\,68\,C, which is exactly co-spatial
with the HI distribution.

The 15\arcsec\ HI beam resolves the gas in DDO\,68\,C, and there is
evidence for rotation in the images shown in Figure~\ref{figcap3}.
The velocity field of DDO\,68\,C was derived using Gaussian fitting to
emission above the 4\,$\sigma$ level via the GIPSY task XGAUFIT.  We
do not have sufficient spatial resolution to attempt a formal
dynamical analysis of DDO\,68\,C.  However, a major-axis
position-velocity (PV) slice through the 15\arcsec\ datacube suggests
a minimum (i.e., uncorrected for inclination, which is likely
significant based on the far-UV morphology) rotation velocity of 7.5-10
\kms.

We estimate the SF rate (SFR) using the GALEX far-UV image shown in
Figure~\ref{figcap3}.  We derive an integrated far-UV magnitude
m$_{\rm FUV}$ $=$ 19.3 for DDO\,68\,C.  Using the prescriptions in
\citet{salim07}, this corresponds to a far-UV SFR (SFR$_{\rm FUV}$) of
(1.4\,$\pm$\,0.4)\,$\times$10$^{-3}$ \msun\,yr$^{-1}$.  This 
SFR$_{\rm FUV}$ is similar to those of the least massive dwarfs in
\citet{lee09}.  For comparison, the SFR$_{\rm FUV}$ of
DDO\,68 is $\sim$20 times larger (SFR$_{\rm FUV} =$ 0.023
\msun\,yr$^{-1}$ and 0.029 \msun\,yr$^{-1}$ from {Lee
  \etal\ 2009}\nocite{lee09} and {Hunter
  \etal\ 2010}\nocite{hunter10}, respectively).

DDO\,68\,C is not detected in the WIYN 0.9m continuum-subtracted
\halpha\ image.  Assuming a point source, the upper limit of the
integrated \halpha\ luminosity is L$_{\rm H\alpha}$ $<$
8.5\,$\times$\,10$^{36}$ erg\,s$^{-1}$, corresponding to an
\halpha-based SFR upper limit SFR$_{\rm H\alpha}$ $<$
7\,$\times$\,10$^{-5}$ \msun\,yr$^{-1}$.  These lines of evidence
suggest that despite the higher SFR$_{\rm FUV}$ within the past few
hundred Myr, the galaxy is now undergoing a period of relative
quiescence or has a stochastically-sampled upper IMF.

%-----------------------------------------------------------------------------%
\section{Evidence for Interaction}
\label{S4}
%-----------------------------------------------------------------------------%

In Figure~\ref{figcap4} we show channel maps of the DDO\,68-DDO\,68\,C
system derived from our GBT OTF mapping observations.  These images
were created by blanking emission below the 3$\sigma$ level in the
smoothed 15.46 km\,s$^{-1}$\,ch$^{-1}$ data cube ($\sigma =$6 mK in
line-free channels).  Across multiple channels, emission extends from
DDO\,68 toward, and overlapping with, DDO\,68\,C.  The two galaxies
are connected by a bridge of low surface brightness gas with
integrated HI column densities $\lsim$5\,$\times$\,10$^{18}$
cm$^{-2}$.

The HI connecting DDO\,68 and DDO\,68\,C is apparent in both the GBT
and the VLA data.  In Figure~\ref{figcap5} we show HI column density
images, Gaussian-fitted velocity fields, and PV slices from both
datasets.  Panels (a) and (b) show moment maps derived from the VLA
data after applying spectral smoothing and Gaussian tapering in the
$uv$ plane; the resulting beam size and velocity resolution are
300\arcsec\ and 10.3 km\,s$^{-1}$\,ch$^{-1}$, respectively. At this
resolution and sensitivity, the low surface brightness HI gas appears
between the two systems at a column density N$_{\rm HI}$
$\simeq$2\,$\times$\,10$^{18}$ cm$^{-2}$.  In panels (d) and (e), the
GBT data (beam size $=$ 523.3\arcsec) clearly show the extension of
the HI gas toward and enclosing DDO\,68\,C.  This extended emission is
apparent at integrated column densities of 1\,$\times$\,10$^{18}$
cm$^{-2}$ $\lsim$ N$_{\rm HI}$ $\lsim$ 10$^{19}$ cm$^{-2}$ in the GBT
data.  Panels (c) and (f) show PV slices taken at a position angle of
355\arcdeg\ (measured counter-clockwise from north) and passing
through the HI surface density maxima of both DDO\,68 and DDO\,68\,C.
These panels clearly verify the presence of low surface brightness HI
gas between DDO\,68 and DDO\,68\,C.

It is important to note that the angular separation of the two
galaxies (11.34\arcmin\ using the coordinates discussed above) is
slightly larger than the synthesized beam of the GBT OTF maps
(8.7\arcmin).  As Figure~\ref{figcap4} shows, a GBT beam resolution
element centered on each source does not overlap with the other;
formally, the two sources are resolved.  The presence of low surface
brightness HI gas in the VLA data assures that this bridge is not
entirely a resolution or smoothing effect.  Very deep, low-resolution
interferometric images of this system would be valuable in further
studying the nature of the gas between DDO\,68 and its companion.

We interpret the low surface brightness HI gas connecting DDO\,68
and its companion as direct evidence for an ongoing interaction.  This
interpretation is strengthened by the various lines of discussion in
\S~\ref{S1}.  Specifically, the optical and HI morphologies of DDO\,68
are both severely disturbed.  Further, the ongoing massive SF is
concentrated in regions of the outer disk and in the stellar stream
extending to the south.

%-----------------------------------------------------------------------------%
\section{Discussion}
\label{S5}
%-----------------------------------------------------------------------------%

The ongoing interaction of DDO\,68 and DDO\,68\,C provides a possible
explanation for the deviation of DDO\,68 from the M-Z relationship.
The present-day metallicity of the HII regions may be a result of the
complex mixing of infalling neutral material.  As discussed in detail
in \citet{sancisi08}, the close proximity of gas-rich companions and
the presence of HI and stellar tails (all of which are seen in the
DDO\,68-DDO\,68\,C system) provide compelling support for ongoing cold
gas accretion.

With the present data we cannot conclude that the interaction is
responsible for infall of pristine gas into the DDO\,68 disk.  While
it is possible that deep, interferometric low-resolution HI imaging
could separate infalling gas from material associated with an
interaction, it is likely that an alternative method will be needed to
identify an infall episode.  UV absorption line spectroscopy offers
one avenue for such investigation \citep[see,
  e.g.,][]{lebouteiller13}.  An alternative method would be to examine
elemental abundances in all available HII regions in the disk; to date
only abundances for the northern HII regions have been derived
\citep{pustilnik05,berg12}.  A measurement of the abundances in the
HII region in the stellar stream to the south (see \halpha\ image in
Figure~\ref{figcap1}) would reveal if the entire disk of DDO\,68 has
experienced a uniform level of chemical enrichment.  Unfortunately,
the \halpha\ non-detection of DDO\,68\,C precludes a spectroscopic
measurement of its chemical abundance.
 
As DDO\,68 and its companion appear to be interacting, it would be 
especially interesting to compare the recent
SF histories (SFHs) of these systems.  As discussed in McQuinn
\etal\ ({2010a}\nocite{mcquinn10a}, {2010b}\nocite{mcquinn10b},
{2012}\nocite{mcquinn12}), spatially-resolved SFHs offer unique
insights into the locations and intensities of SF as functions
of time.  New HST observations of DDO\,68\,C that are similar
to the archival observations of DDO\,68 shown in
Figure~\ref{figcap1} would allow a comparative SFH analysis
between these two galaxies.  The \halpha\ non-detection but
significant far-UV luminosity of DDO\,68\,C enforces the
importance of probing the evolution of these systems over the
past few hundred Myr.  Correlated features in these SFHs would
provide an empirical timescale on the interaction event.

%-----------------------------------------------------------------------------%
\acknowledgements
%-----------------------------------------------------------------------------%

J.M.C. is supported by NSF grant 1211683.  The authors thank
D.J. Pisano, Spencer Wolfe, and Dominic Ludovici for sharing their
software, and Glen Langston for help with imaging the GBT data.  Some
of the results in this paper are based on observations made with the
NASA/ESA Hubble Space Telescope, obtained from the Data Archive at the
Space Telescope Science Institute, which is operated by the
Association of Universities for Research in Astronomy, Inc., under
NASA contract NAS 5-26555. The HST observations presented in this work
are associated with program \#11578.

%-----------------------------------------------------------------------------%
%\clearpage
\bibliographystyle{apj}                                                 

%-----------------------------------------------------------------------------%

\end{document}